%% file: MnTiO3.tex
\documentclass[prb,onecolumn,superscriptaddress,floatfix,showpacs,amsmath,amssymb]{revtex4-2}

\usepackage{graphicx}
\usepackage{dcolumn,xcolor}
\usepackage{bm}
\usepackage{color}
\usepackage{tabularx}
\usepackage{siunitx}
\usepackage{verbatim}
\usepackage{parskip}
\usepackage{soul}
\usepackage[normalem]{ulem}
\usepackage{setspace}

\newcommand{\mto}{MnTiO$_3$}
\newcommand{\cto}{CoTiO$_3$}
\newcommand{\nto}{NiTiO$_3$}

\newcommand{\tn}{$T_{\rm N}$}
\newcommand{\mb}{$\mu_{\rm B}$}
\newcommand{\mbfu}{$\mu_\mathrm{B}$/f.u.}
\newcommand{\cpx}{$c_{\rm p}$}

\newcommand{\cpph}{$c_{\rm p}^{\rm ph}$}

\newcommand{\jmk}{J/(mol\,K)}

\newcommand{\bsf}{$B_{\rm{SF}}$}
\newcommand{\bsfesr}{$B^{\rm ESR}_{\rm{SF}}$}

\newcommand{\ali}{$\alpha_{\rm i}$}

\newcommand{\alb}{$\alpha_{\rm b}$}
\newcommand{\alc}{$\alpha_{\rm c}$}

\newcommand{\ba}{$B_\mathrm{A}$}
\newcommand{\be}{$B_\mathrm{E}$}

\newcommand{\oup}{$\nu_{||}^{\rm {U}}$}
\newcommand{\odw}{$\nu_{||}^{\rm {L}}$}
\newcommand{\osf}{$\nu_{||}^{\rm {SF}}$}
\newcommand{\operp}{$\nu_{\perp}$}

\usepackage{xr}
\makeatletter
\newcommand*{\addFileDependency}[1]{
  \typeout{(#1)}
  \@addtofilelist{#1}
  \IfFileExists{#1}{}{\typeout{No file #1.}}
}
\makeatother

\newcommand*{\myexternaldocument}[1]{
    \externaldocument{#1}
    \addFileDependency{#1.tex}
    \addFileDependency{#1.aux}
}
\myexternaldocument{MnTiO3-SI}
\begin{document}

\title{The role of magnetoelastic coupling and magnetic anisotropy in MnTiO$_3$}

\author{L.~Gries}
\altaffiliation{Both authors contributed equally to this work.}
\affiliation{Kirchhoff Institute of Physics, Heidelberg University, INF 227, D-69120 Heidelberg, Germany}

\author{M.~Jonak}
\altaffiliation{Both authors contributed equally to this work.}
\affiliation{Kirchhoff Institute of Physics, Heidelberg University, INF 227, D-69120 Heidelberg, Germany}

\author{A.~Elghandour}
\affiliation{Kirchhoff Institute of Physics, Heidelberg University, INF 227, D-69120 Heidelberg, Germany}
\author{K.~Dey}
\altaffiliation{Current address: Clarendon Laboratory, University of Oxford, Parks Road, Oxford, OX1 3PU, United Kingdom}

\affiliation{Kirchhoff Institute of Physics, Heidelberg University, INF 227, D-69120 Heidelberg, Germany}

\author{R.~Klingeler}
\email{klingeler@kip.uni-heidelberg.de}\affiliation{Kirchhoff Institute of Physics, Heidelberg University, INF 227, D-69120 Heidelberg, Germany}

\date{\today}

\begin{abstract}

We report the thermodynamic properties studied by thermal expansion, magnetostriction, magnetisation, and specific heat measurements as well as the low-energy magnetic excitations of \mto\ and investigate how magneto-elastic coupling and magnetic anisotropy affect the evolution of long-range order and the magnetic phase diagram. Specifically, we utilise high-resolution capacitance dilatometry and antiferromagnetic resonance (AFMR) studies by means of high-frequency electron spin resonance (HF-ESR) spectroscopy. The role of anisotropy is reflected by spin-reorientation at \bsf~$\simeq 6$~T and a corresponding sign change in $\partial T_{\rm N}/\partial B$. Analysis of the AFMR modes enables us to establish the zero-field excitation gap $\Delta$ as well as its temperature dependence. We derive the effective anisotropy field $B_\mathrm{A} = 0.16(1)$~T which predominately originates from out-of-plane nearest-neighbour dipole-dipole interactions. Despite the nearly fully quenched orbital moment, our data show pronounced thermal expansion and magnetostriction anomalies at \tn\ and \bsf , respectively, which allows the experimental determination of sizable uniaxial pressure dependencies, i.e., $\partial B_{\rm SF}/\partial p_{\rm c} = -0.20(2)$~T/GPa, $\partial T_{\rm N}/ \partial p_b = 0.69(12)$~K/GPa, and $\partial T_{\rm N}/ \partial p_c = -2.0(4)$~K/GPa. Notably, short-range magnetic order appears up to at least 3\tn , as indicated by anisotropic lattice distortion, the violation of a constant Grüneisen behavior, and the presence of local magnetic fields detected by HF-ESR. 

\end{abstract}

\maketitle

\section{Introduction}

Ilmenite-structured titanates $M$TiO$_3$ ($M$ = Mn, Fe, Ni, Co) have been studied for nearly a century, with the observation of significant magnetoelectric effects in CoTiO$_3$, NiTiO$_3$ \cite{Harada.2016} and \mto\ \cite{Mufti.2011}, placing them into the class of multiferroics and reviving interest in fundamental research on them. Besides potential applications in multiferroic technology, the ilmenite titanates show promising applications as photocatalysts \cite{Truong.2012, Wang.2016, Wang.2018, Shu.2008}, in solar cells \cite{Wang.2016b, Li.2018}, and as sensors \cite{He.2008, Lu.2015, Aparna.2019}. Furthermore, interesting effects, such as NiTiO$_3$ thin films showing polarization-induced ferromagnetism \cite{Varga.2017} and the existence of Dirac magnons in \cto , have been reported recently \cite{Yuan.2020}.

In  \mto , G-type long-range antiferromagnetic order evolves at \tn\ = 64~K and a linear magnetoelectric effect is observed as demonstrated by anomalies in the dielectric function $\epsilon$ for $B > 0$~T in the ordered phase and the appearance of finite electric polarisation~\cite{Akimitsu.1970, Akimitsu.1977, Shirane.1959, Stickler.1967,Mufti.2011, Chen.2014}. The material crystallizes in the space group $R\bar 3$ where 2/3 inhabited hexagonal layers of Mn$^{2+}$ and Ti$^{4+}$ ions surrounded by distorted oxygen octahedra are stacked~\cite{Barth.1934}. Notably, the magnetic susceptibility shows a broad maximum around $100~$K which is attributed to in-plane two-dimensional short-range order persisting at least up to 2\tn . This is supposed to result from accidental cancellation of the interlayer exchange interactions~\cite{Akimitsu.1970,Akimitsu.1977, Todate.1986}. Below \tn , uniaxial anisotropy favors the easy magnetic axis along the crystallographic $c$ direction and a spin-flop transition appears at 
$B_{\rm SF} || c = 6~$T~\cite{Yamauchi.1983, Stephenson.1968}. It has been recently shown that spin-reorientation is accompanied by a flop in electric polarization $P$ from the $c$-axis to the $a$-axis~\cite{Silverstein.2016}.

Here, we report the interplay of spin and lattice in \mto , and the role of small but finite magnetic anisotropy. We apply high-resolution capacitance dilatometry to obtain thermal expansion and magnetostriction up to 15~T on \mto\ single crystals which are supported by specific heat and magnetisation data. The thermodynamic response functions are combined with antiferromagnetic resonance studies of the $q=0$ magnon excitations. Our experimental data enable us to quantify and elucidate the magnetoelastic coupling and the role of magnetic anisotropy. In addition, the magnetic phase diagram is established.

\section{Experimental Methods}

Single crystals of \mto\ were grown using the floating-zone technique in a high-pressure optical furnace (HKZ, SciDre) equipped with a 3500-W Xe arc lamp~\cite{Neef.2017}. The precursor for crystal growth was prepared via a standard solid-state reaction of stoichiometric amounts of MnCO$_3$ and TiO$_2$, performed at \ang{1200}C in air. Macroscopic single crystals were grown in the argon atmosphere (flow rate 0.5~l/min) at an elevated pressure of 5~bar. The growth rate was maintained at 5~mm/h and the feed and seed rods were counter-rotated at 10~rpm. Phase purity of the resulting single crystals was confirmed using powder X-ray diffraction on a Bruker D8 Advance ECO diffractometer. Laue diffraction shows high crystallinity and was used to orient the single crystal. For the measurements reported at hand, a cuboid-shaped single crystal of dimensions of about $1.2 \times 2.1 \times 2.1~$mm$^{3}$ was used [see Fig.~\ref{fig:Crystal_Pic} in the Supplemental Information (SI)].

The magnetisation was investigated in the temperature regime 2--300~K and in magnetic fields up to 14~T using a vibrating sample magnetometer (VSM) of Quantum Design's Physical Property Measurement System (PPMS-14). The specific heat was studied in the same device using a relaxation method. The relative length changes $dL_i/L_i$ were studied by means of a three-terminal high-resolution capacitance dilatometer in a home-built setup~\cite{Kuchler.2017, Werner.2017}, and the thermal expansion coefficients $\alpha_i = 1/L_i \times dL_i(T)/dT$ were obtained. The magnetostriction, i.e., the field-induced length change $dL_i(B)/L_i$, was measured at several fixed temperatures in magnetic fields up to 15~T and the longitudinal magnetostriction coefficients $\lambda_i = 1/L_i\times dL_i(B)/dB$ were derived. In all dilatometric studies, the magnetic field was applied along the direction of the measured length changes. High-frequency electron spin resonance studies were performed in the frequency range 30~GHz $< \nu <$ 420~GHz utilising a Millimetre-Wave Vector Network Analyzer from AB Millimetre~\cite{Comba.2015}. Moreover, X-Band frequency measurements were performed on a Bruker Elexsys E500 spectrometer. While all the above-mentioned studies were performed on the same single crystal, auxiliary HF-ESR measurements were conducted also on a finely-ground polycrystalline sample fixed by means of eicosane. 

\section{Results}

\subsection{Evolution of antiferromagnetic order and magneto-structural coupling at $B=0$~T}

\begin{figure}
    \centering
    \includegraphics[width=0.9\columnwidth,clip]{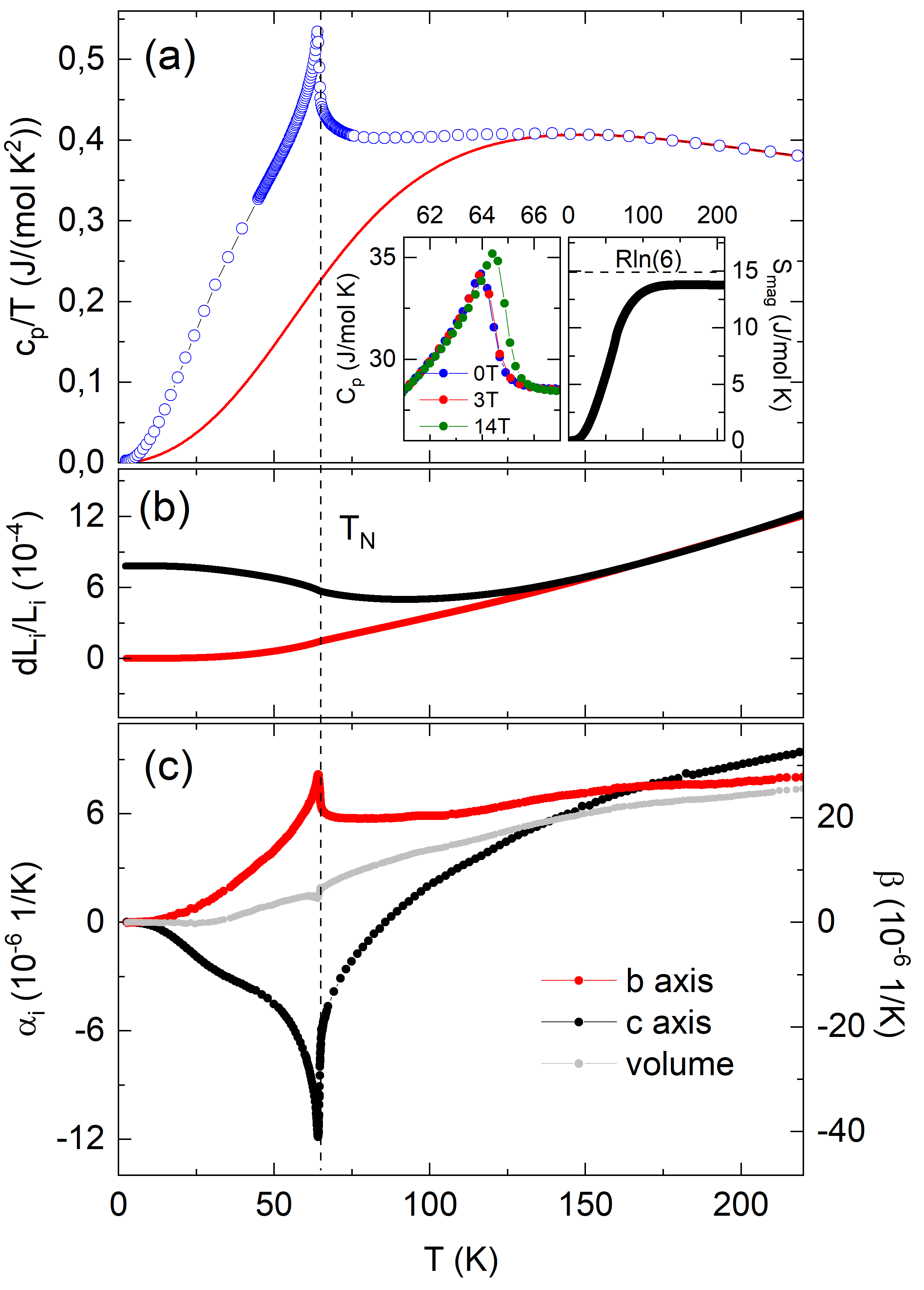}
    \caption{(a) Temperature dependence of the specific heat \cpx $/T$ in zero magnetic field. The solid red line indicates the phonon specific heat \cpph\ (see the text). The vertical dashed line marks \tn. Insets: Specific heat \cpx\ at selected applied magnetic fields $B ||  c$ axis around the Néel temperature and magnetic entropy obtained by integrating $(c_{\rm p} - c_{\rm p}^{\rm ph}) / T$. (b) Relative length changes $dL_i/L_i$ along the crystallographic $b$ and $c$ axes $vs.$ temperature. The data have been shifted vertically such, that they overlap at $200$~K.  (c) Corresponding thermal expansion coefficients \ali .}
    \label{fig:TE_CP}
\end{figure}

The specific heat displayed as \cpx$/T$ in Fig.~\ref{fig:TE_CP}(a) shows a $\lambda$-like anomaly at $T_{\rm N}=64(1)$~K which signals a second order phase transition associated with the evolution of long-range antiferromagnetic order in MnTiO$_3$. This is in accordance with previous studies on \mto\ \cite{Stickler.1967, Stephenson.1968, Akimitsu.1970}. Corresponding clear features in the relative length changes $dL_i/L_i$ and the thermal expansion coefficient $\alpha_i$, i.e., a kink and a $\lambda$-shaped peak, respectively, for the crystallographic directions $i=b,c$ (Fig.~\ref{fig:TE_CP}(b),(c)) show the existence of pronounced magneto-elastic coupling in MnTiO$_3$. Differing signs of the anomalies in $\alpha_i$ imply opposite uniaxial pressure dependencies of \tn , specifically a positive dependence for pressure applied along the $b$ axis and negative for $p||c$ axis. The volume thermal expansion, approximated by $\beta= (\alpha_c + 2 \cdot \alpha_b)$, shows only a small (negative) anomaly in its temperature derivative, corresponding to negative hydrostatic pressure dependence of \tn .

In order to obtain the total magnetic entropy changes $S_{\rm mag}(T)$, we approximate the phononic specific heat \cpph\ with an Einstein-Debye model in a temperature regime well above the Néel temperature (i.e., above 150~K). The fit yields the characteristic temperatures $\Theta_{\rm D} = 508$~K and $\Theta_{\rm E} = 1254$~K and describes the data well for temperature above $140$~K (see the red line in Fig.~\ref{fig:TE_CP}(a)). Integrating $(c_{\rm p} - c_{\rm p}^{\rm ph}) / T$ yields the magnetic entropy changes (right inset of Fig.~\ref{fig:TE_CP}(a)), which at $T>140$~K nearly agree to the full expected magnetic entropy $S_{\rm mag}^{\rm th} = R \ln{6}=14.9$~J/(mol K) associated with the $S = 5/2$ spin system. At \tn , only 2/3 of the total magnetic entropy is released, i.e., there are pronounced magnetic entropy changes above \tn . This finding confirms the presence of significant short-range order above \tn\ which has been suggested to be of two-dimensional nature~\cite{Akimitsu.1970, Akimitsu.1977, Todate.1986}. 

\begin{figure}[htb]
    \centering
    \includegraphics[width=0.9\columnwidth,clip]{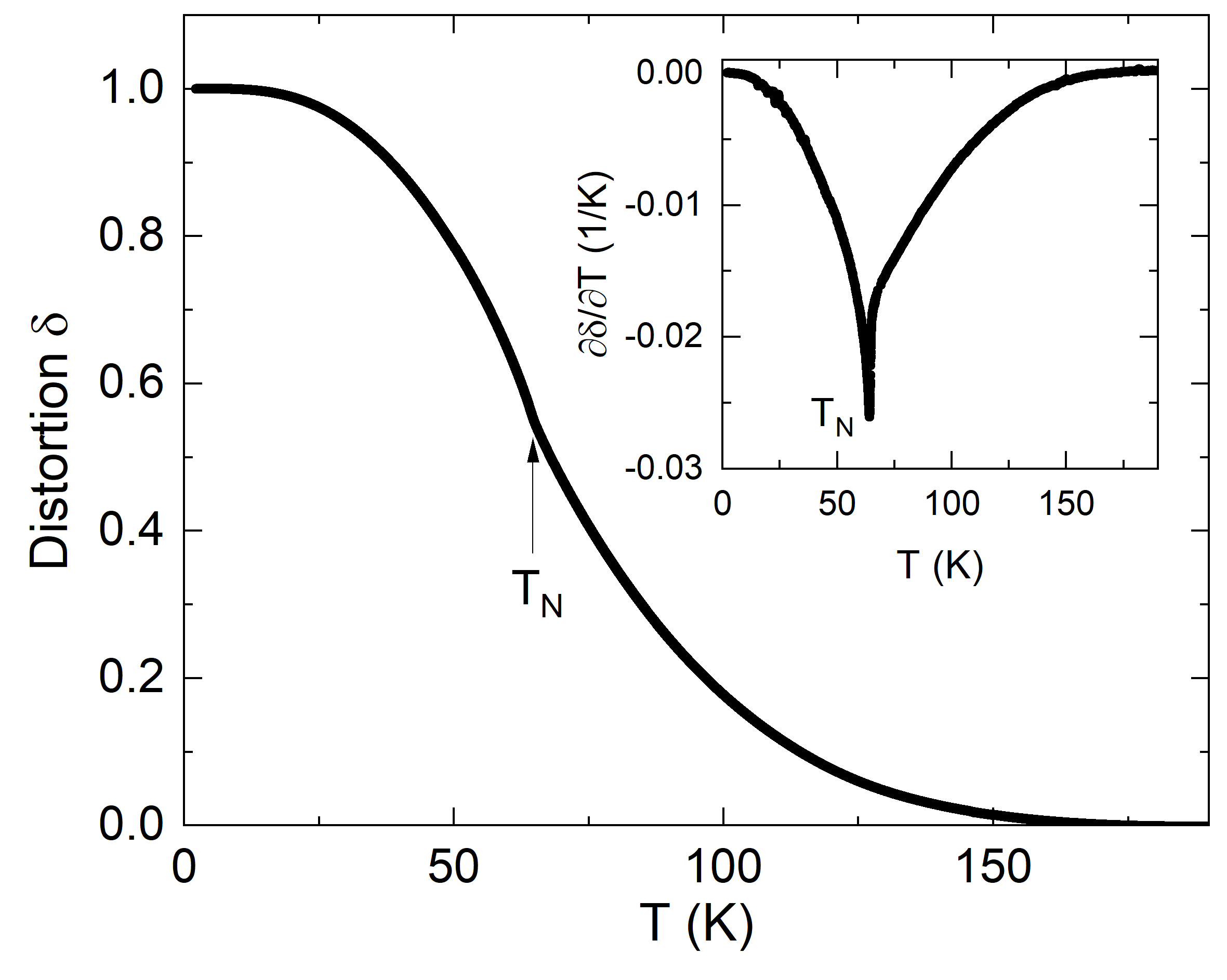}
    \caption{Distortion parameter $\delta = (dL_\mathrm{c} - dL_\mathrm{b})/(dL_\mathrm{c} + dL_\mathrm{b})$ as a function of temperature. Inset: Temperature derivative of the distortion parameter $\partial \delta / \partial T$.}
    \label{fig:distortion}
\end{figure}

Thermal expansion of the $b$ and $c$ axis is very similar at high temperatures as shown in Fig.~\ref{fig:TE_CP}(b) but becomes anisotropic below $\thicksim170$~K. While the $b$ axis shrinks upon cooling in the whole temperature regime under study, the $c$ axis displays a minimum around $T\simeq 85$~K below which it expands towards lowest temperatures. This is reflected by the sign of the thermal expansion coefficients with \alb~$>0$ and a sign change in \alc\ at 85~K (Fig.~\ref{fig:TE_CP}(c)). In addition to the anomalies at \tn , \alc\ exhibits a broad minimum around $30$~K which is not observed in \alb . The anisotropy in thermal expansion is described by a distortion parameter $\delta = (dL_\mathrm{c} - dL_\mathrm{b})/(dL_\mathrm{c} + dL_\mathrm{b})$ shown in Fig.~\ref{fig:distortion}. Upon cooling, $\delta$ evolves finite values and gradually increases below $\thicksim170$~K. At low temperatures, structural distortion saturates and becomes almost constant below $20$~K. At \tn , $\delta$ exhibits a moderate anomaly which appears $\lambda$-shaped in the temperature derivative $\partial\delta /\partial T$ (see the inset of Fig.~\ref{fig:distortion}). It is straightforward to attribute the observed distortion to magneto-elastic coupling and spin degrees of freedom driving anisotropic thermal expansion both, in the long-range (below \tn ) and short-range spin ordered regimes.

\begin{figure}[htb]
    \centering
    \includegraphics[width=0.9\columnwidth,clip]{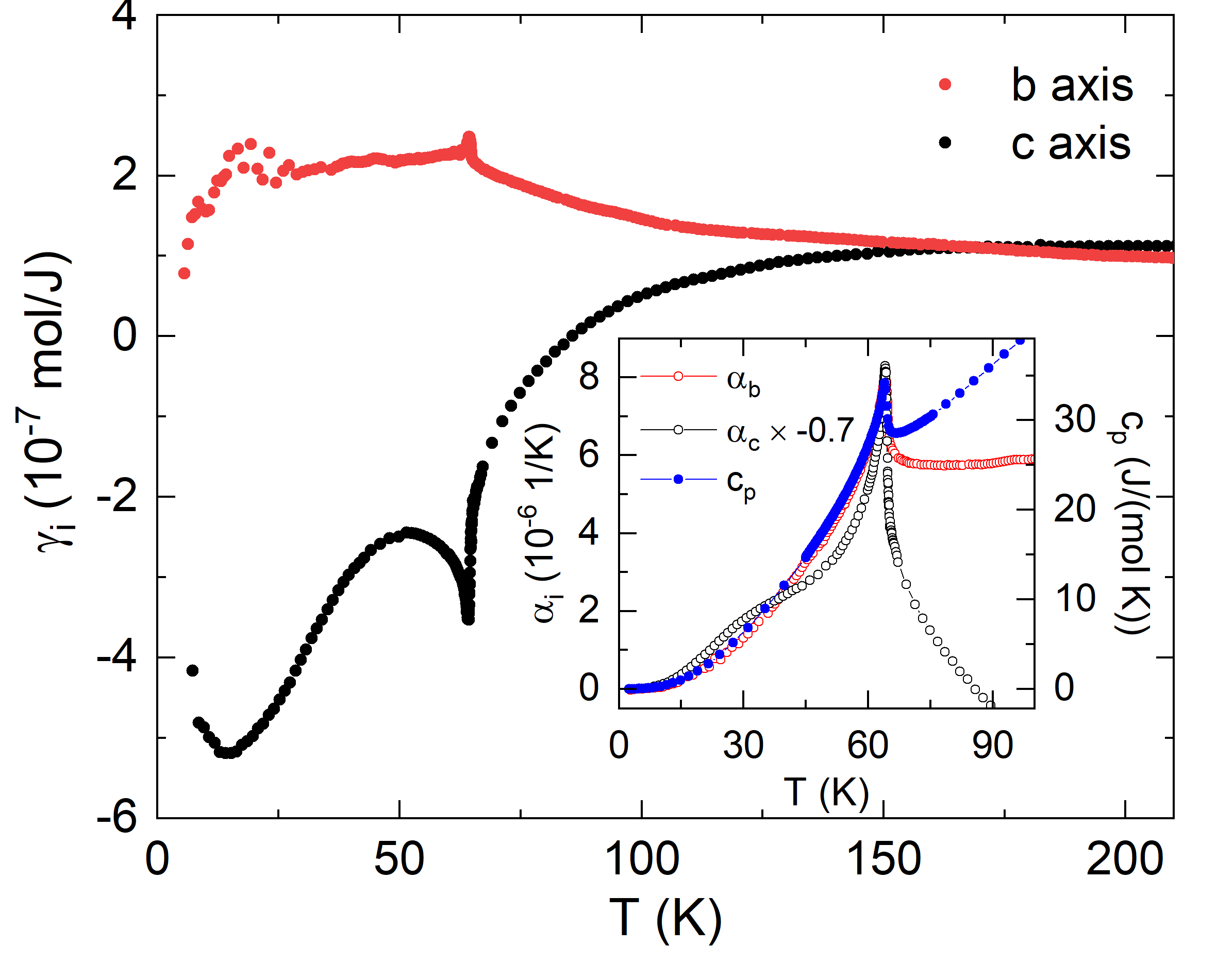}
    \caption{Uniaxial Grüneisen ratio $\gamma_i = \alpha_i/c_{\rm p}$ for $i = b$ and $c$ axis. Inset: Thermal expansion coefficients $\alpha_i$ (left ordinate) and specific heat \cpx\ (right ordinate) scaled such that the anomalies at \tn\ overlap, with $\alpha_c$ being multiplied with $-0.7$.}
    \label{fig:gruen}
\end{figure}

Further information on the evolution of the low-temperature phase is obtained by comparing the entropy and the lattice changes through the (uniaxial) Grüneisen parameter $\gamma_i =$~\cpx /\ali . Here, we consider the experimentally obtained total thermal expansion and specific heat, since our data do not allow us to extract the purely magnetic length changes because a reliable phonon background could  not  be obtained. The uniaxial Grüneisen parameters (see Fig.~\ref{fig:gruen}) are approximately constant at high temperatures, thereby indicating a single dominating energy scale~\cite{Klingeler.2006, Gegenwart.2016}. This agrees with the fact that entropy changes in this temperature regime are attributed to phonons (see Fig.~\ref{fig:TE_CP}a). The corresponding Grüneisen ratio amounts to $\gamma = 10.5(9) \times 10^{-7}$~mol/J. The relevance of an additional (magnetic) degree of freedom below $\thicksim 170$~K yields a non-constant behaviour for both axes with opposite signs. For example, $\gamma_{\rm b}$ increases towards the onset of long-range antiferromagnetic order and exhibits an anomaly at the transition, followed by a nearly temperature-independent regime down to $\thicksim 22$~K. In contrast, $\gamma_{\rm c}$ exhibits a more pronounced anomaly at \tn\ and varies more strongly upon cooling. We conclude that the lattice changes along the $b$ axis are dominated by one magnetic energy scale at 20~K $\lesssim T <$ \tn , while the behavior of the $c$ axis is driven by more than one degree of freedom. This is further illustrated in the inset of Fig.~\ref{fig:gruen}, where \alc\ shows a generally different behavior below \tn\ as compared to \cpx , while \alb\ can be scaled to \cpx\ by a single factor.

By exploiting the Ehrenfest relation $dT_{\rm N}/dp_{\rm i} = T_{\rm N} V_m \Delta \alpha_{\rm i} / \Delta c_{\rm p}$, the uniaxial pressure dependencies of \tn\ can be derived from the anomalies. In order to acquire the anomaly heights in \cpx\ and \ali , a method of equal-area construction~\cite{Kuchler.2005} was applied (see Fig.~\ref{fig:SAC} in the SI). For \mto\ this method leads to $\Delta c_{\rm p} = 9.0(16)$~\jmk , $\Delta \alpha_{\rm b} = 3.0(5) \times 10^{-6}$~1/K and $\Delta \alpha_{\rm c} = -7.5(13) \times 10^{-6}$~1/K. Considering the molar volume $V_m = 3.2 \times 10^{-5}$~m$^3$/mol yields the uniaxial pressure dependencies $\partial T_{\rm N}/ \partial p_b = 0.69(12)$~K/GPa and $\partial T_{\rm N}/ \partial p_c = -2.0(4)$~K/GPa for uniaxial pressure along the $b$ and $c$ axis, respectively.

\subsection{Magnetostriction and spin-reorientation}

\begin{figure}[htb]
    \centering
    \includegraphics[width=0.9\columnwidth,clip]{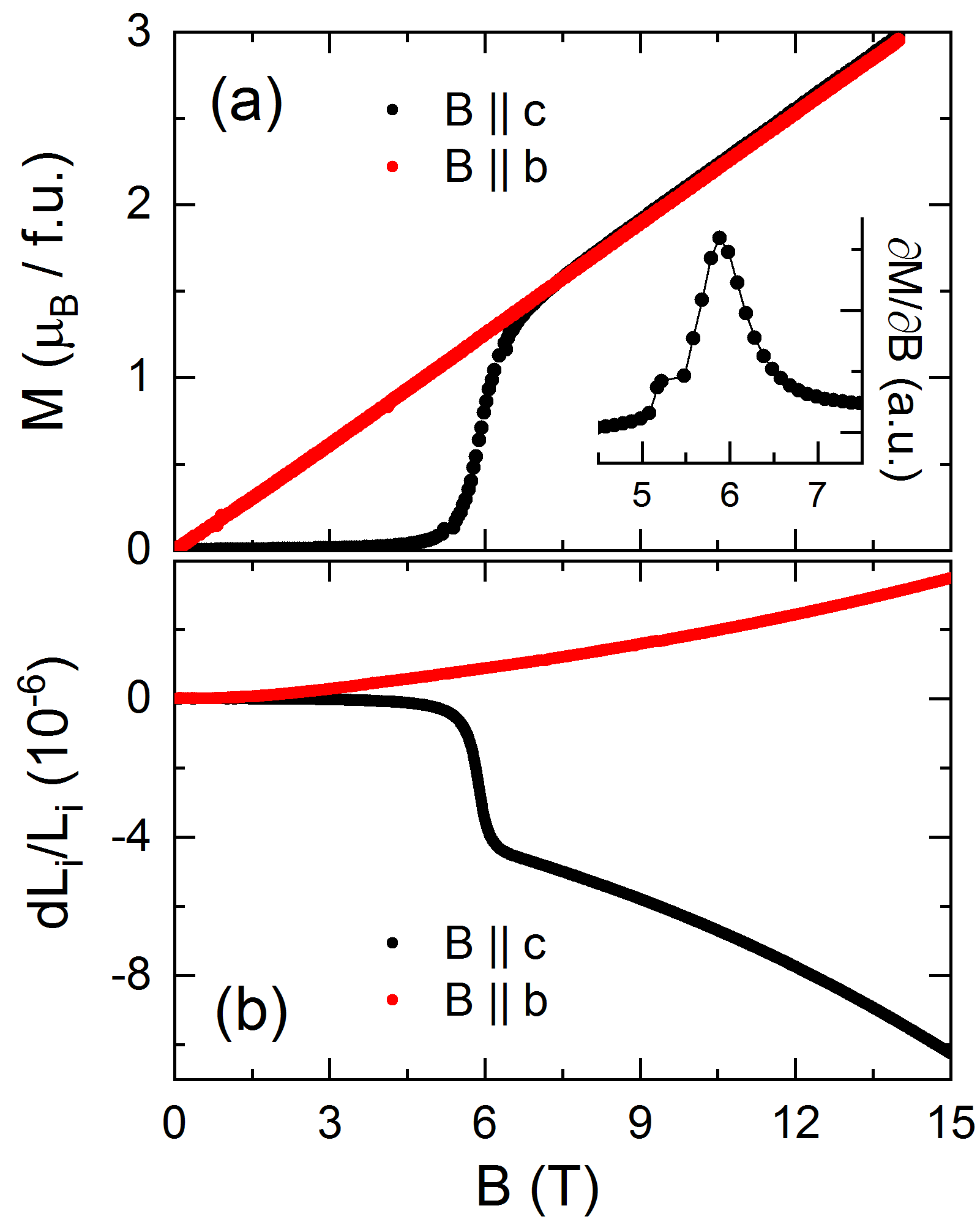}
    \caption{(a) Magnetisation $M(B)$ and (b) magnetostriction $dL_i(B)/L_i$ for $B || b$ and $B || c$ axis, respectively, at 2~K. Inset: Corresponding isothermal susceptibility $\partial M/\partial B$ for $B || c$. 
    }
    \label{fig:M(B)}
\end{figure}

The isothermal magnetisation $M$ at $2$~K for magnetic fields up to $14$~T applied along the $b$ and $c$ axis is plotted in Fig.~\ref{fig:M(B)}(a). For $B || b$ axis, the magnetisation shows a linear behaviour with $\chi_\mathrm{b}=\partial M/\partial B= 0.021(1)$~$\mu_B$/(f.u.\,T). In contrast, for $B || c$ a jump in $M$ is observed at \bsf~$=5.9(1)$~T, separating linear behaviour in $M$ vs.~$B$ with $\chi_\mathrm{c}\simeq \chi_\mathrm{b}$ in high fields from $\chi_\mathrm{c}\simeq 0$ for $B<$~\bsf . These observations are typical for a spin-flop transition where the spins rotate from the magnetically easy $c$ axis into the $ab$ plane. From the magnetisation jump at \bsf, which amounts to $\Delta M = 0.12$~\mbfu, and from the ordered magnetic moment of 4.55~\mbfu~\cite{Shirane.1959} we infer that for $B\gtrsim B_{\rm SF}$ the spins are predominately in the $ab$ plane with only a small canting of $\simeq \ang{1.5}$ towards the $c$ axis. 
Notably, the spin-reorientation is associated with a discontinuous decrease of the $c$ axis (Fig.~\ref{fig:M(B)}(b)), too, which confirms the relevance of spin-orbit coupling in \mto\ and shows that the transition is not a pure spin phenomenon. We also note that in the linear regimes in $M$ vs.~$B$, where the spins are  approximately perpendicular to $B$, there is a quadratic behaviour of lattice parameters $b$ and $c$. Specifically, for $B || c>$~\bsf , the $c$ axis shrinks quadratically in $B$, while for $B || b$ the $b$ axis elongates up to highest fields.

\begin{figure}[htb]
    \centering
    \includegraphics[width=1\columnwidth,clip]{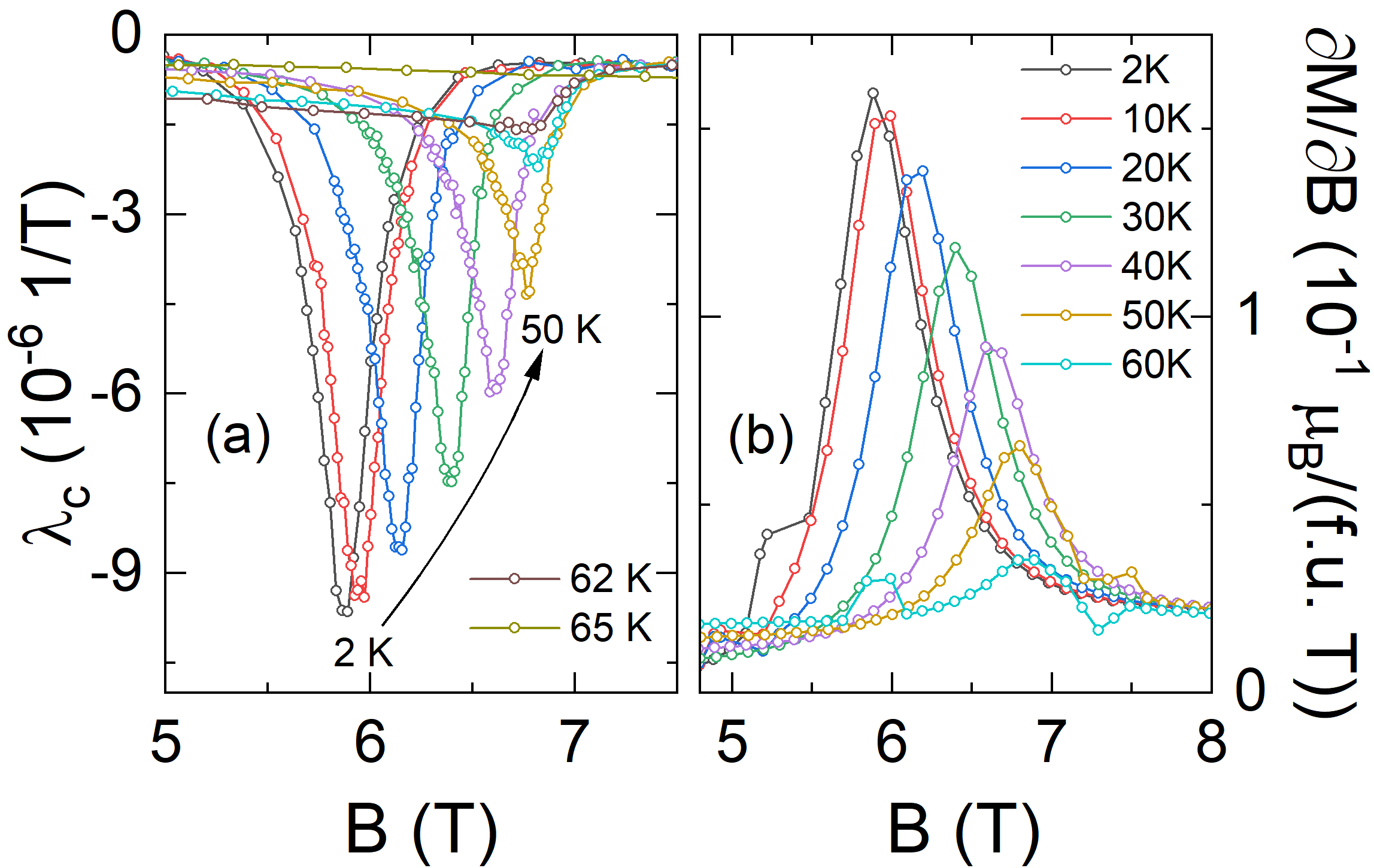}
 \caption{(a) Magnetostriction coefficient $\lambda_c$ and (b) isothermal susceptibility $\chi_c=\partial M/\partial B$ at various temperatures for $B || c$ axis.}
    \label{fig:lambda_c}
\end{figure}

The temperature dependence of the spin-flop-related anomalies driven by $B || c$ is shown in Fig.~\ref{fig:lambda_c} which displays the magnetostriction coefficient $\lambda_c$ and the isothermal susceptibility $\chi_c$. Upon heating, the anomalies shift to higher fields and become smaller, i.e., the associated jumps in magnetisation and length decrease. At 60~K, just below \tn , we find $B_{\rm SF}=6.8(1)$~T. Above the Néel temperature no anomalies are observed.

The fact that the lattice responds to spin-reorientation implies that uniaxial pressure can be used to shift the associated phase boundary. Qualitatively, the data show that uniaxial pressure $p||c$ axis initially yields a decrease of \bsf . For $p_{\rm c}\rightarrow 0$, the uniaxial pressure dependence is associated with the jumps $\Delta M_{\rm c}$ and $\Delta L_{\rm c}$ in magnetisation and length through the Clausius-Clapeyron relation $\partial B_{\rm SF}/\partial p_{\rm c} = V_m/L_c \times \Delta L_c/ \Delta M_c$. At $T=2$~K, the jumps amount to $\Delta M_{\rm c} = 0.12(1)$~$\mu_{\rm B}$/f.u. and $\Delta L_c/L_c = -4.2(4) \times 10^{-6}$, which leads to a uniaxial pressure dependence for $p || c$ of $\partial B_{\rm SF}/\partial p_{\rm c} = -0.20(2)$~T/GPa. For higher temperatures, this value only changes slightly and at $T=40$~K we find $\partial B_{\rm SF}/\partial p_c = -0.16(2)$~T/GPa.

\subsection{Antiferromagnetic resonance}

ESR studies are used to investigate the paramagnetic response as well as the evolution of local magnetic fields and the long-range ordered phase. While in a paramagnetic regime ESR measurements provide information, e.g., on the evolution of short-range magnetic order, in the long-range antiferromagnetically ordered phase HF-ESR is susceptible to collective $q=0$ spin excitations, i.e.~it detects AFMR modes. 

\begin{figure}
\centering
\includegraphics[width=1\columnwidth,clip]{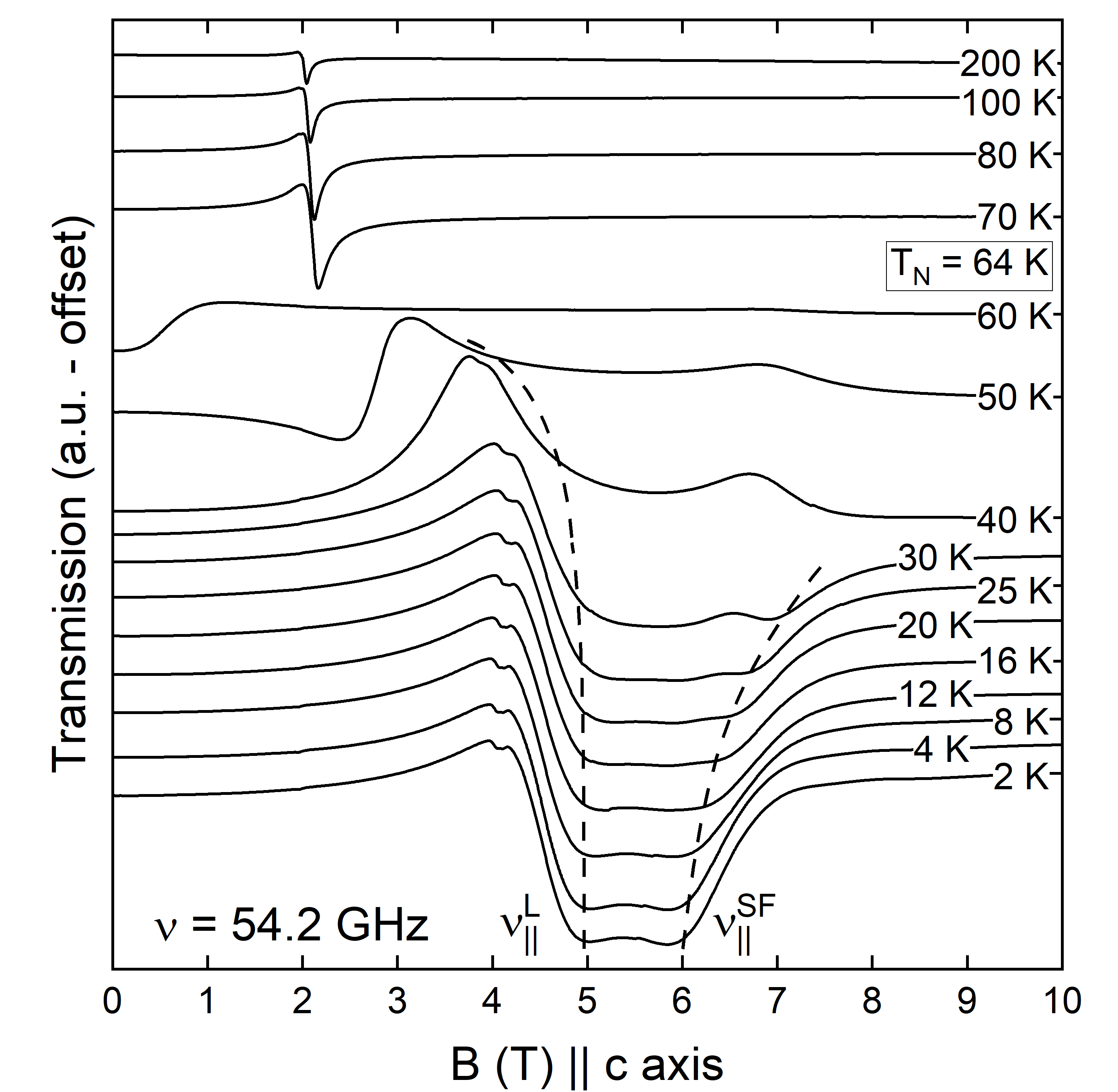}
\caption{Temperature evolution of the resonance spectra recorded at fixed frequency $\nu = 54.2$~GHz and $B || c$ axis. \odw\ and \osf\ mark the lower parallel branch and the spin-flop branch, respectively (see the text). Curved dashed lines are a guide to the eye illustrating the evolution of the resonance features.}
\label{fig:ESR_temp}
\end{figure}

The room temperature X-Band data demonstrate $g_{\rm || c}=2.004(1)$ and $g_{\rm \perp c}=2.002(1)$, as expected for paramagnetic, uncorrelated Mn$^{2+}$ ions with $S = 5/2$ and vanishing orbital momentum (see Fig.~\ref{fig:X-band_RT} in the Supplemental Information)~\cite{Khomskii.2014}. Decreasing the temperature results in a shift of the resonance field away from the $g\simeq 2.00$ resonance position, as attested by our high-frequency measurements. In particular, the resonance spectrum obtained at $T=200$~K and $\nu=54.2$~GHz on the single-crystal sample with $B || c$ axis exhibits a sharp paramagnetic Lorentzian resonance signal with a small effect of wave-phase mixing and with corresponding $g$-factor $g=1.94(4)$ (see Fig.~\ref{fig:ESR_temp}). The deviation from the $g=2.00$ resonance position at $T=200$~K indicates the presence of short-range magnetic correlations at temperatures as high as $T \simeq 3T_\mathrm{N}$. Upon further cooling, the paramagnetic resonance feature broadens and shifts to higher magnetic fields, which clearly demonstrates a further development of the short-range magnetic order. Below \tn , the single paramagnetic resonance feature is found to split, drastically broaden and shift, indicating the evolution of static internal fields. At 2~K, two distinct AFMR modes, marked \odw\ and \osf\ in Fig.~\ref{fig:ESR_temp}, are observed to lie at about 5~T and 6~T, respectively.

\begin{figure}
\centering
\includegraphics[width=1\columnwidth,clip]{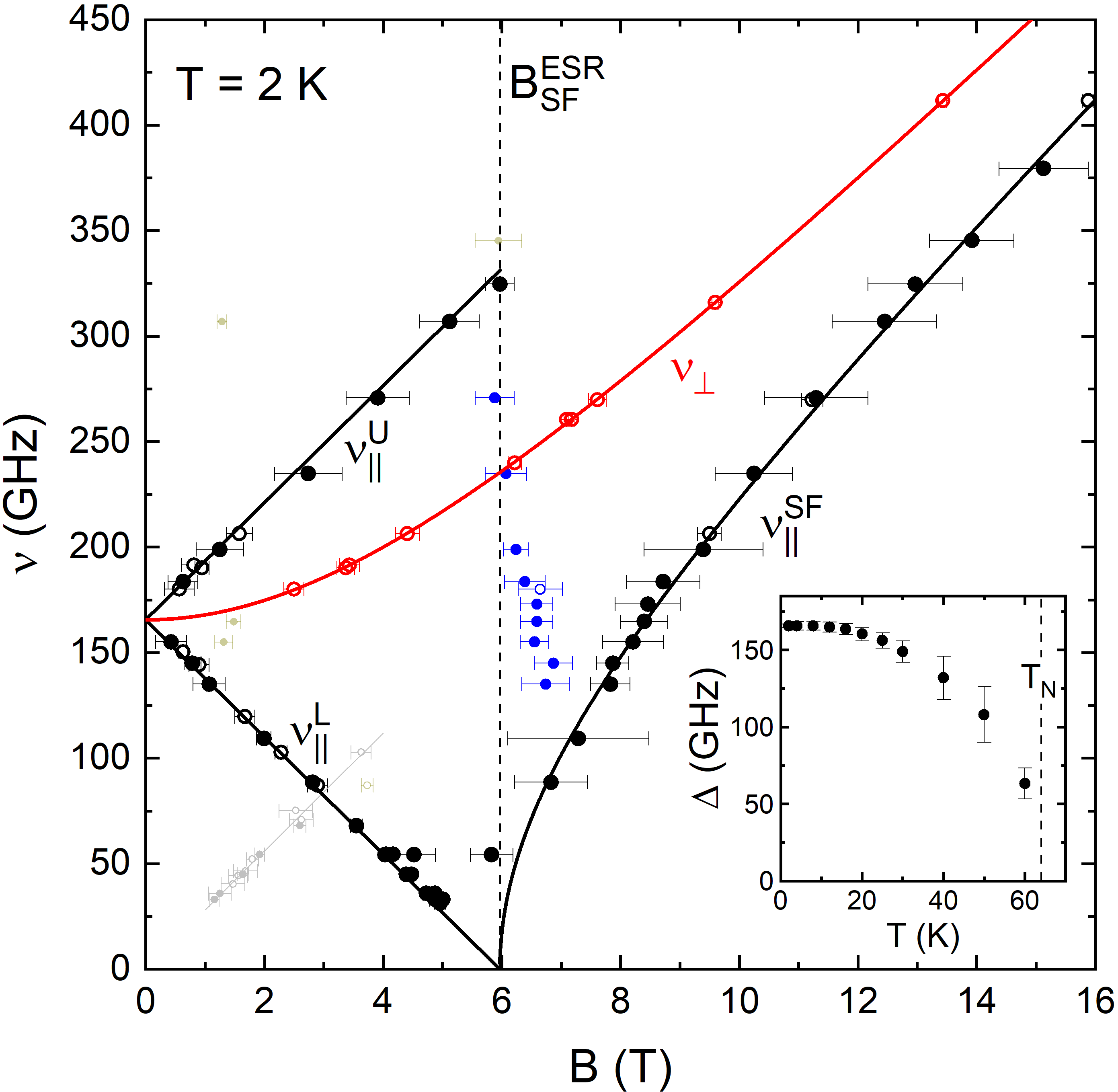}
\caption{Microwave frequencies vs.~resonance magnetic fields at $T=2$~K. Closed symbols depict data obtained on the single crystal with $B || c$ axis, open symbols on the polycrystal. Black and red solid lines show fitting results from a two-sublattice uniaxial AFMR model (see the text). Light grey markers and line corresponds to a weak paramagnetic impurity feature. The vertical dashed line marks the spin-flop field \bsfesr\ as determined from the fitting procedure of the resonance modes (see the text). Inset: temperature dependence of the zero-field excitation gap $\Delta$, whereby the vertical dashed line marks \tn.}
\label{fig:ESR_phd}
\end{figure}

HF-ESR measurements at $T=2$~K and various fixed frequencies allow us to construct the frequency--magnetic-field diagram (Fig.~\ref{fig:ESR_phd}). The figure combines resonance positions stemming from single-crystal measurements with $B || c$ axis and from measurements on the polycrystalline sample. In addition to the dominant resonance features (black and red data points in Fig.~\ref{fig:ESR_phd}), we find weak features forming a gapless branch with $g = 2.00$ and disappearing for $\nu \gtrsim 90$~GHz (see Fig.~\ref{fig:ESR_phd}) which we attribute to a small number of paramagnetic impurities and do not consider for further analysis.

The presence of four dominant resonance modes, three of which merge into a single zero-field excitation gap, indicates a two-sublattice antiferromagnet (AFM) with an axial-like anisotropy. The corresponding model's free parameters are given by the zero-field excitation gap $\Delta$ and by a pair of effective $g$-factors, $g_{|| \mathrm{easy}}$ and $g_{\perp \mathrm{easy}}$~\cite{Keffer.1952, Rado.1963, Rezende.2019}. A least-squares fitting of the four magnon branches yields $\Delta = 166(1)$~GHz, $g_{|| \mathrm{easy}} = 1.98(1)$, and $g_{\perp \mathrm{easy}} = 2.00(1)$. In the two-sublattice uniaxial AFMR model, the spin-flop field is fixed at $B_\mathrm{SF}^{\rm ESR} = (h\Delta) / (g_{|| \mathrm{easy}}\mu_\mathrm{B})$, where $h$ is the Planck constant and $\mu_\mathrm{B}$ the Bohr magneton, giving $B_\mathrm{SF}^{\rm ESR} = 6.0(1)$~T. Note that using an isotropic $g$-factor as suggested by high-temperature X-band ESR data does not lead to a difference in the determined value of the gap. The hereby-obtained value of the spin-flop field corroborates its determination from the magnetisation and magnetostriction measurements presented in Fig.~\ref{fig:M(B)} and is in accordance with previous studies on \mto\ \cite{Yamauchi.1983, Steiner.1996}. The fitting results of the magnon branches are displayed in Fig.~\ref{fig:ESR_phd} as black and red solid lines, and the resonance branches are labelled \oup, \odw, \osf, and \operp , corresponding to the upper, lower, and spin-flop branch stemming from $B~||$ easy axis, and to the perpendicular branch stemming from $B\perp$ easy axis, respectively. Further to the dominant strongly field-dependent features, almost field-independent resonance features around the position of the spin-flop field have also been detected for $B || c$ (blue data points in Fig.~\ref{fig:ESR_phd}). Such a field-independent resonance branch is expected to occur at \bsf , corresponding to the rotation of the spins perpendicularly to the easy axis~\cite{Rezende.2019}.

The AFMR results confirm the orientation of the easy axis along the crystallographic $c$ axis, as the single-crystal measurements with $B || c$ axis contribute solely to the model's predictions for the easy-axis resonance features. The model's above-obtained parameters contrast with earlier observations from HF-ESR in which the zero-field excitation gap was found to amount to $\Delta = 153$~GHz and the effective $g$-factor to 2.1 \cite{Stickler.1962}, and to $\Delta = 156$~GHz \cite{Stickler.1967}, respectively; and from inelastic neutron scattering which found $\Delta \simeq 200$~GHz~\cite{Todate.1986}, and $\Delta \simeq 193$~GHz~\cite{Hwang.2021}, respectively. The high quality of the single crystal at hand and the superior energy resolution of ESR lend support to the present values of $\Delta$ and  $g$-factors in MnTiO$_3$.

Having shown that the two resonance features \odw\ and \osf\ can be interpreted as the lower and spin-flop branch of a uniaxial two-sublattice AFMR model, the features' temperature dependence can be revisited. As observed in Fig.~\ref{fig:ESR_temp}, \odw\ remains at approximately the same resonance field all the way up to 30~K. For $T \geq 40$~K, a significant shift of the resonance field to lower values is detected. Since in the model under consideration, \odw\ is directly related to the zero-field splitting, the branch's shift to lower fields signals reduction in $\Delta$ and its eventual disappearance as $T\nearrow T_\mathrm{N}$. Indeed, making use of~\cite{Keffer.1952}

\begin{equation}
    \nu_{||}^\mathrm{L} = \left[\Delta^2 + \left(\frac{\chi_{||}}{2\chi_\perp}\frac{g_{||}\mu_\mathrm{B}B}{h}\right)^2\right]^{\frac{1}{2}}-\frac{g_{||}\mu_\mathrm{B}B}{h}\left(1-\frac{\chi_{||}}{2\chi_\perp}\right)
\label{eq:AFMR_lower_branch}
\end{equation}

where $\chi_{||}$ ($\chi_\perp$) is the temperature-dependent static magnetic susceptibility parallel (perpendicular) to the easy magnetisation axis (Fig.~\ref{fig:static_magnetic_susceptibility} in the SI), the temperature evolution of the feature's position may be used to obtain the temperature dependence of $\Delta$. This is displayed in the inset of Fig.~\ref{fig:ESR_phd} and exhibits a  typical behaviour of a long-range-ordered AFM. A similar temperature evolution of the zero-field excitation gap was observed also in previous HF-ESR measurements on \mto~\cite{Stickler.1962,Stickler.1967}.

The resonance branch \osf\ is found to shift to higher resonance fields as temperature increases from the lowest measured temperatures, and, moreover, no resonance feature associated with \osf\ is observed for $T \geq 40$~K. A similar effect has been observed in a number of long-range-ordered AFMs, such as CuCl$_2\cdot2$H$_2$O~\cite{Nagamiya.1954} or LiFePO$_4$~\cite{Werner.2021}. Note that the conventional mean-field spin-only AFMR model predicts a shift of \osf\ to lower resonance fields upon heating, contrary to our observation.

\section{Discussion}

\begin{figure}[htb]
    \centering
    \includegraphics[width=1\columnwidth,clip]{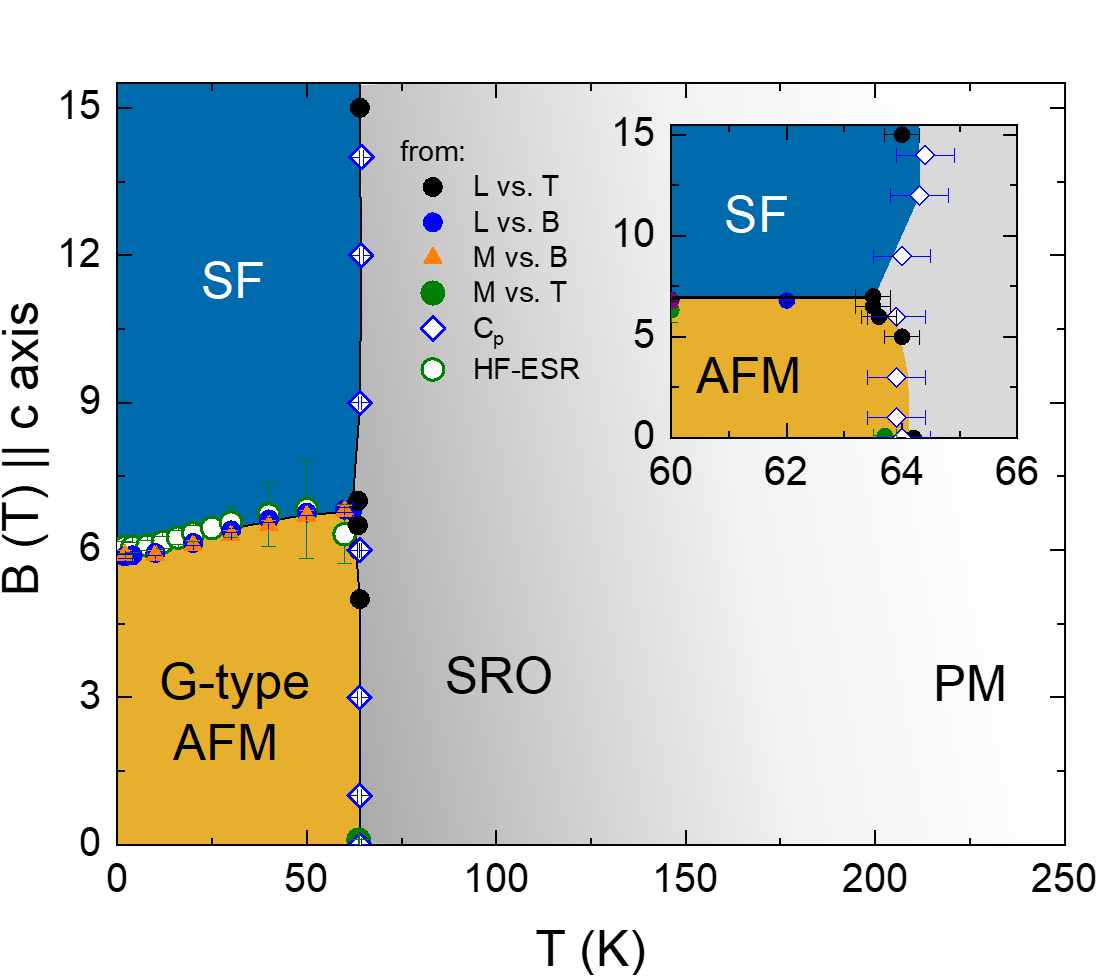}
    \caption{Magnetic phase diagram of \mto\ for $B||c$ axis constructed from magnetisation $M(T,B)$, dilatometry $L(T,B)$, specific heat $C_{\rm p}(T,B)$, and HF-ESR data. Lines are guides to the eye. G-type AFM, SF, PM, and SRO label the G-type antiferromagnetic, the spin-flop, the high-temperature paramagnetic, and the paramagnetic short-range-ordered phase, respectively. The inset enlarges the regime around \tn . }
    \label{fig:PD_c}
\end{figure}

Our thermodynamic studies presented in Figs.~\ref{fig:TE_CP}, \ref{fig:M(B)}, \ref{fig:lambda_c}, and \ref{fig:alpha_in_field} allow us to construct the magnetic phase diagram in Fig.~\ref{fig:PD_c}. 
Also included in the phase diagram is $B_\mathrm{SF}$ as calculated from the temperature-dependent HF-ESR spectra (Fig.~\ref{fig:ESR_temp}) by means of Eq.~\ref{eq:AFMR_lower_branch}. 
The results show that application of $B || c <$~\bsf\ yields a small reduction of the long-range ordering temperature \tn\ by about 0.5~K. In contrast, $\partial T_{\rm N}/\partial B_{\rm ||c}$ is positive for fields $B>$~\bsf\ and \tn\ increases to 65~K at $15$~T. Thermodynamically, according to the Ehrenfest relation this observation implies that the phase boundary \tn ($B>$~\bsf ) is associated with an increase in magnetic susceptibility upon cooling. In addition, as discussed above, \bsf\ is found to increase slightly upon heating (see Figs.~\ref{fig:lambda_c}). Specifically, it increases by $\simeq 0.9$~T from $5.9(1)$~T at $2$~K to $6.8(1)$~T at $62$~K. For fields applied along the crystallographic $b$ axis, the N\'{e}el temperature does not significantly change up to the highest measured fields. 

The calculated uniaxial pressure dependencies of \tn\ show that in-plane pressure stabilizes antiferromagnetic order, while the order is suppressed for $p || c$. Based on inelastic neutron scattering the net in-plane antiferromagnetic exchange interaction $J_{\rm ab}$ is much stronger than the net out-of-plane interaction $J_{\rm c}$, with ratio of $J_{\rm c}/J_{\rm ab}=0.022$ at $2$~K~\cite{Hwang.2021}. Upon heating, this ratio further diminishes to less than $1\%$ at \tn~\cite{Hwang.2021}. Assuming that long-range order depends on $J_{\rm c}$ much more sensitively than on  $J_{\rm ab}$, the experimentally determined uniaxial pressure dependencies of \tn\ reflect predominately $\partial J_{\rm c}/\partial p_{\rm i}$. This is consistent with our results which indicate opposite signs of the uniaxial pressure dependencies at \tn, as uniaxial pressure along $c$ will decrease the interplanar distances, while $p||ab$ in a simple macroscopic picture is expected to have an opposite effect.

A comparison of $\partial T_{\rm N}/\partial p_{\rm i}$ in \mto\ with values from the literature for \nto~\cite{Dey.2020} and \cto~\cite{Hoffmann.2021} shows that despite strong differences in spin-orbit coupling and despite different types of antiferromagnetic ordering, the uniaxial pressure dependencies have the same order of magnitude. This again is consistent with the supposition of the dominant role of $J_{\rm c}$ for establishing long-range magnetic order in these materials. Note, however, that the three magnetic titanates have different signs of the uniaxial pressure dependence at \tn: $\partial T_{\rm N}/\partial p_{\rm c}$ is negative in \nto\ and \cto, while it is positive in \mto; for
$p||ab$, the uniaxial pressure dependence is positive in \mto\ and \cto, while it is negative in \nto\ (see Table \ref{tab:comparison_titanates}).

\begin{table}
\centering
\begin{tabular}{c|c|c|c}
\hline\hline

   & MnTiO$_3$ & NiTiO$_3$ & CoTiO$_3$\\
\hline
    $\partial T_N / \partial p_{\rm \perp c}$ & + & - & + \\
    $\partial T_N / \partial p_{\rm \| c}$ & - & + & + \\
    spin~structure & G-type ($|| c$)  & A-type ($\perp c$) & A-type ($\perp c$)\\

\hline
\hline

\end{tabular}
\caption{Signs of the uniaxial pressure dependencies $\partial T_{\rm N}/\partial p_{\rm i}$ of \tn\ for \mto , \nto~\cite{Dey.2020} and \cto~\cite{Hoffmann.2021}, as well as the antiferromagnetic spin configuration in the respective material.}
    \label{tab:comparison_titanates}
\end{table}

In addition to these quantitative results describing the spin-lattice coupling in \mto , our data allow conclusions about the coupling of spin and lattice to the dielectric properties. Previous measurements show that \mto\ develops spontaneous polarisation in $c$ direction below \tn\ in finite fields $B||c > 0$ which increases with field up to $B_{\rm SF}$ and is approximately constant in temperature in the ordered phase~\cite{Silverstein.2016, Chi.2014}. At $B=0$~T, no spontaneous polarisation is observed. The lattice, on the other hand, exhibits significant anisotropic changes with temperature below roughly $170$~K in zero magnetic field, as can be seen from the distortion parameter in Fig.~\ref{fig:distortion} as well as in finite magnetic fields (see Fig.~\ref{fig:alpha_in_field}). Furthermore, applying magnetic field parallel to the $c$ axis does not induce relevant lattice changes below $B_{\rm SF}$ as shown in Fig.~\ref{fig:M(B)} and Fig.~\ref{fig:alpha_in_field}. This implies that the strong magnetostrictive effects which we observe at $B=0$~T do not drive the ferroelectric moment but additional, i.e.~field-induced, symmetry breaking is essential.

By performing a linear extrapolation of the static magnetisation in Fig.~\ref{fig:M(B)}(a) to the known value of the full ordered magnetic moment in \mto ,  $\mu_\mathrm{sat}=4.55$~\mbfu~\cite{Shirane.1959}, the saturation field can be roughly estimated as $B_\mathrm{sat}=213(10)$~T. In combination with the spin-flop field $B_\mathrm{SF}=5.9$~T observed in our study (cf.~Figs.~\ref{fig:M(B)} and \ref{fig:ESR_phd}), the effective anisotropy and exchange fields, \ba\ and \be , can be obtained. To do so, we take the spin Hamiltonian~\cite{Rezende.2019}

\begin{equation}
\hat{\mathcal{H}} = - D\sum_{i}(\hat{S}_i^\mathrm{z})^2 + \sum_{<i,j>}J_{ij}\hat{\bm{S}}_i\cdot \hat{\bm{S}}_j - g\mu_{\mathrm B}\sum_{i}\bm{B}\cdot \hat{\bm{S}}_i
\label{eq:spin_Hamiltonian}
\end{equation}

where the first term quantifies the anisotropy energy, the second term the exchange energy, and the third term the Zeeman energy. $D$ is the uniaxial anisotropy constant, defined as $D=(g\mu_\mathrm{B}B_\mathrm{A})/(2S)$~\cite{Rezende.2019}. $J_{ij}$ is the magnetic exchange constant between two magnetic neighbours occupying the sites $i$ and $j$.

To solve the Hamiltonian in Eq.~\ref{eq:spin_Hamiltonian}, we assume a two-sublattice model and relate macroscopic sublattice magnetisations to the above spin operators via the relation $\bm{M}_{1,2}=g\mu_\mathrm{B}N\hat{S}_{i,j}$, where $N$ is the number of spins per unit volume~\cite{Rezende.2019}. Considering the exchange interaction only between nearest neighbours residing on opposite sublattices, the effective exchange interaction can be defined as $J_\mathrm{eff}=(g\mu_\mathrm{B}B_\mathrm{E})/(2Sz)$, where $z$ is the number of nearest neighbours~\cite{Rezende.2019}. Upon minimisation of the energy in Eq.~\ref{eq:spin_Hamiltonian}, the expressions $B_\mathrm{A}=B_\mathrm{SF}^2/B_\mathrm{sat}$, $B_\mathrm{E}=B_\mathrm{sat}/2 + B_\mathrm{SF}^2/2B_\mathrm{sat}$, and $\cos(\theta)=B/(2B_\mathrm{E}-B_\mathrm{A})$ are obtained in terms of experimentally measurable quantities, where $\theta$ is the angle between the easy axis and the sublattice magnetisation in the spin-flop phase. We find \ba~$=0.16(1)$~T and \be~$=107(6)$~T, which correspond to $D=0.0038(3)$~meV and, with the number of nearest neighbours $z=3$, $J_\mathrm{eff}=0.82(4)$~meV. In addition, $\theta_\mathrm{SF}=88.4(1)$\textdegree\ at the spin flop. 

The signs of the obtained parameters from Eq.~\ref{eq:spin_Hamiltonian} signal that the system's anisotropy assumes a uniaxial character and that the dominant intersublattice exchange interaction is antiferromagnetic. Moreover, the obtained values of $D$ and $J_\mathrm{eff}$ quantitatively corroborate their determination by means of inelastic neutron scattering: $D=0.0011$~meV and $J_1=0.63$~meV~\cite{Todate.1986}, and $D=0.0045$~meV and $J_1=0.92$~meV~\cite{Hwang.2021}, respectively~\footnote{$J_1$ denotes the dominant, nearest-neighbour exchange coupling; as such, $J_\mathrm{eff}$ in the present study reflects predominately $J_1$.}. A good correspondence is found also between the angle $\theta_\mathrm{SF}$ as estimated from the size of $\Delta M$ at the spin flop transition (cf.~Fig.~\ref{fig:M(B)}(a)) and as ascertained here from the energy minimisation of the above Hamiltonian. 

In a mean-field model, the effective anisotropy field can be independently calculated if the perpendicular static magnetic susceptibility, $\chi_\perp$, is known~\cite{Barak.1978}:

\begin{equation}
B_\mathrm{A} = \frac{B_\mathrm{SF}^2 \cdot \chi_\perp}{M_\mathrm{sat}}
\end{equation}

Our static magnetic susceptibility obtained for $B || b$ axis reveals that $\chi_\mathrm{b}=0.0122(2)$~erg/(G$^{2}$mol) at 2~K (see Fig.~\ref{fig:static_magnetic_susceptibility} in the SI). Hence, the effective anisotropy field can be estimated as $B_\mathrm{A}=0.17(1)$~T, corroborating the above determination of its value.

The ratio of the anisotropy to exchange field amounts to $B_\mathrm{A}/B_\mathrm{E} \simeq 1.5 \times 10^{-3}$, thereby confirming the weak effective anisotropy field relative to the effective exchange field. This ratio is typical for Mn$^{2+}$-containing two-sublattice AFMs such as P$\bar{3}$1$m$-phase MnSb$_2$O$_6$, P321-MnSb$_2$O$_6$, CaMnCl$_3$ $\cdot$ H$_2$O, or MnF$_2$ which have been studied by AFMR~\cite{Koo.2018, Werner.2016, Phaff.1983, Barak.1978}. In MnF$_{2}$, $B_\mathrm{A}=0.82$~T and $B_\mathrm{E}=52.6$~T, resulting in the ratio $B_\mathrm{A}/B_\mathrm{E} \simeq 1.6 \times 10^{-2}$~\cite{Barak.1978} while the other examples feature values of $B_\mathrm{A}/B_\mathrm{E} \simeq 2\times 10^{-3}$.

In searching for the origin of the anisotropy, it can be observed that due to vanishing orbital moment of the Mn$^{2+}$ magnetic ions, spin-orbit coupling is expected to act only in second order of perturbation theory~\cite{Khomskii.2014}. In particular, only negligible contributions to the overall anisotropy are expected to arise from single-ion effects and from anisotropic exchange interaction. In principle, both symmetric and antisymmetric exchange can lead to macroscopic anisotropy effects~\cite{Khomskii.2014}. Based on symmetry considerations, however, the contribution from antisymmetric exchange in \mto\ amounts to zero, $J_\mathrm{anis}^\mathrm{antisym} = 0$. The contribution due to symmetric exchange can be estimated as $J_\mathrm{anis}^\mathrm{sym}\sim J_\mathrm{eff}(\frac{\Delta g}{g})^2$~\cite{Khomskii.2014}. With the room-temperature spectroscopic $g$-factors $g_{\rm || c}=2.004(2)$ and $g_{\rm \perp c}=2.002(2)$ obtained in our X-Band measurement (cf.~Fig.~\ref{fig:X-band_RT}), $\Delta g = 0.002$, resulting in the effective anisotropy energy on the order of $\sim S^2 J_\mathrm{anis}^\mathrm{sym}\simeq 5 \times 10^{-6}$~meV. This is almost three orders of magnitude smaller than the calculated anisotropy constant $D$ and hence cannot account for its value.

On the other hand, the dipole-dipole energy can be approximated as $E_\mathrm{dip} = \frac{\mu_0 \mu_\mathrm{B}^2 g_1 g_2 }{4\pi |\bm{r}|^3} [3 ( \bm{S}_1 \cdot \bm{\hat{r}} )( \bm{S}_2 \cdot \bm{\hat{r}} ) - \bm{S}_1 \cdot \bm{S}_2]$ where $g_i$ and $\bm{S}_i$ denote the $g$-factor and the spin of the i$^\mathrm{th}$ magnetic moment, and where $\bm{r}$ is the direction vector between two neighbouring magnetic moments. Due to the dot product between the spin orientation and the direction vector, the dipole-dipole interaction between the in-plane nearest-neighbour magnetic moments vanishes to first order. Instead, the dominant contribution to $E_\mathrm{dip}$ stems from interactions between the out-of-plane nearest-neighbour magnetic moments. With the separation of $r=4.01$~\r{A} obtained from our single-crystal crystallographic refinement and using the ordered moment of 4.55~\mb, one finds $E_\mathrm{dip}\simeq 0.034$~meV, i.e.~$\simeq 0.017$~meV per magnetic ion. Although this value is more than four times larger than the above-obtained value of $D$, its proximity to the experimental observations indicates the significance of dipole-dipole interactions for the evolution of anisotropy in \mto . It may be speculated that the discrepancy to the experimental observations results from second-order effects due to in-plane dipole-dipole interactions or from admixing of Mn$^{3+}$ ($3d^4$) ions~\cite{Yamauchi.1983}, both of which may reduce the actual dipole-dipole energy.
We also note that a superexchange mechanism involving the simultaneous transfer of an electron-hole pair between neighbouring ions may be relevant for explaining the observed discrepancy between the experimentally observed anisotropy and the estimated dipolar anisotropy~\cite{Goodenough.1967}. The exciton superexchange will take place along the Mn--O--Mn path in the $ab$-planes and may be therefore expected to reduce the dominant out-of-plane dipole-dipole-caused anisotropy.

While the low-temperature AFMR modes are explained well by the mean-field spin-only model elaborated above, the temperature dependence of the resonance mode \osf\ is not accounted for by the conventional mean-field theory. This predicts a shift of \osf\ to lower resonance fields upon heating, in contrast to our observation of the resonance's shift to higher fields (see Fig.~\ref{fig:ESR_temp}). Furthermore, our results show that, despite acting only to the second order in perturbation theory, spin-orbit coupling must be also considered to fully cover the magnetic ground state properties in \mto. This is clearly evidenced by the observed spin-lattice coupling and magnetostrictive response on spin-reorientation.

\section{Summary}

In summary, we report on the static and dynamic magnetic properties and magneto-elastic coupling in \mto\ single crystals in magnetic fields up to 15~T. We present the magnetic phase diagram which exhibits the spin-flop phase and show that $\partial T_{\rm N}/\partial B$ changes sign at \bsf . 
Despite the vanishing orbital moment associated with Mn$^{2+}$ ions in \mto , we observe pronounced thermal expansion and magnetostriction anomalies at \tn\ and \bsf , respectively, of similar magnitude as in related Co$^{2+}$- and Ni$^{2+}$-based magnetic titanates. Accordingly, uniaxial pressure affects the respective phase boundaries and we obtain a finite suppression of \bsf\ amounting to $\partial B_{\rm SF}/\partial p_{\rm c} = -0.20(2)$~T/GPa. Long-range antiferromagnetic order is stabilised (suppressed) by $p||b$ ($p||c$), as $\partial T_{\rm N}/ \partial p_{\rm b} = 0.69(12)$~K/GPa and $\partial T_{\rm N}/ \partial p_{\rm c} = -2.0(4)$~K/GPa, respectively. The observed AFMR modes are fully described by a two-sublattice mean-field model with uniaxial anisotropy. The model's optimised parameters are the effective $g$-factors $g_{|| \mathrm{easy}} = 1.98(1)$ and $g_{\perp \mathrm{easy}} = 2.00(1)$, and the zero-field excitation gap $\Delta = 166(1)$~GHz. $\Delta$ is found to decrease with increasing temperature and to vanish at \tn. Our analysis yields the effective anisotropy field $B_\mathrm{A} = 0.16(1)$~T, corresponding to $D=0.0038(3)$~meV, as well as $B_\mathrm{A}/B_\mathrm{E}\simeq 1.5 \times 10^{-3}$. The anisotropy is argued to originate predominately from out-of-plane nearest-neighbour dipole-dipole interactions. Notably, short-range magnetic order persists well above \tn , i.e.~up to 3\tn , as indicated by a finite lattice distortion parameter $\delta$, the violation of a constant Grüneisen behavior, and the presence of local magnetic fields up to at least 200~K detected by HF-ESR. Our results hence elucidate the origin and the role of anisotropy and show its relevance as well as the importance of spin-lattice coupling for static and dynamic magnetic properties in \mto.

\section*{Acknowledgements}

We acknowledge financial support by BMBF via the project SpinFun (13XP5088) and by Deutsche Forschungsgemeinschaft (DFG) under Germany’s Excellence Strategy EXC2181/1-390900948 (the Heidelberg STRUCTURES Excellence Cluster) and through project KL 1824/13-1. L.G.~and K.D.~acknowledge support by IMPRS-QD. A.E.~acknowledges support by the DAAD through the GSSP program.

\bibliography{Literature}
 
\clearpage
\newpage

\input{MnTiO3-SI}

\end{document}

%% file: MnTiO3-SI.tex
\setstretch{1.1}
\section*{\large Supplemental Information: The role of magnetoelastic coupling and magnetic anisotropy in MnTiO$_3$}
\setstretch{1}

\hfill
\\
The Supplemental Information presents a photographic image of the crystal studied in this work, it displays supporting static magnetic susceptibility and illustrates how the equal-area construction was used on the specific heat data to estimate phase-transition heights. Furthermore, the Supplemental Information contains additional thermal expansion data in applied magnetic fields and presents room-temperature X-band electron paramagnetic resonance data.

\begin{figure}[h]
    \centering
    \includegraphics[width=0.9\columnwidth,clip]{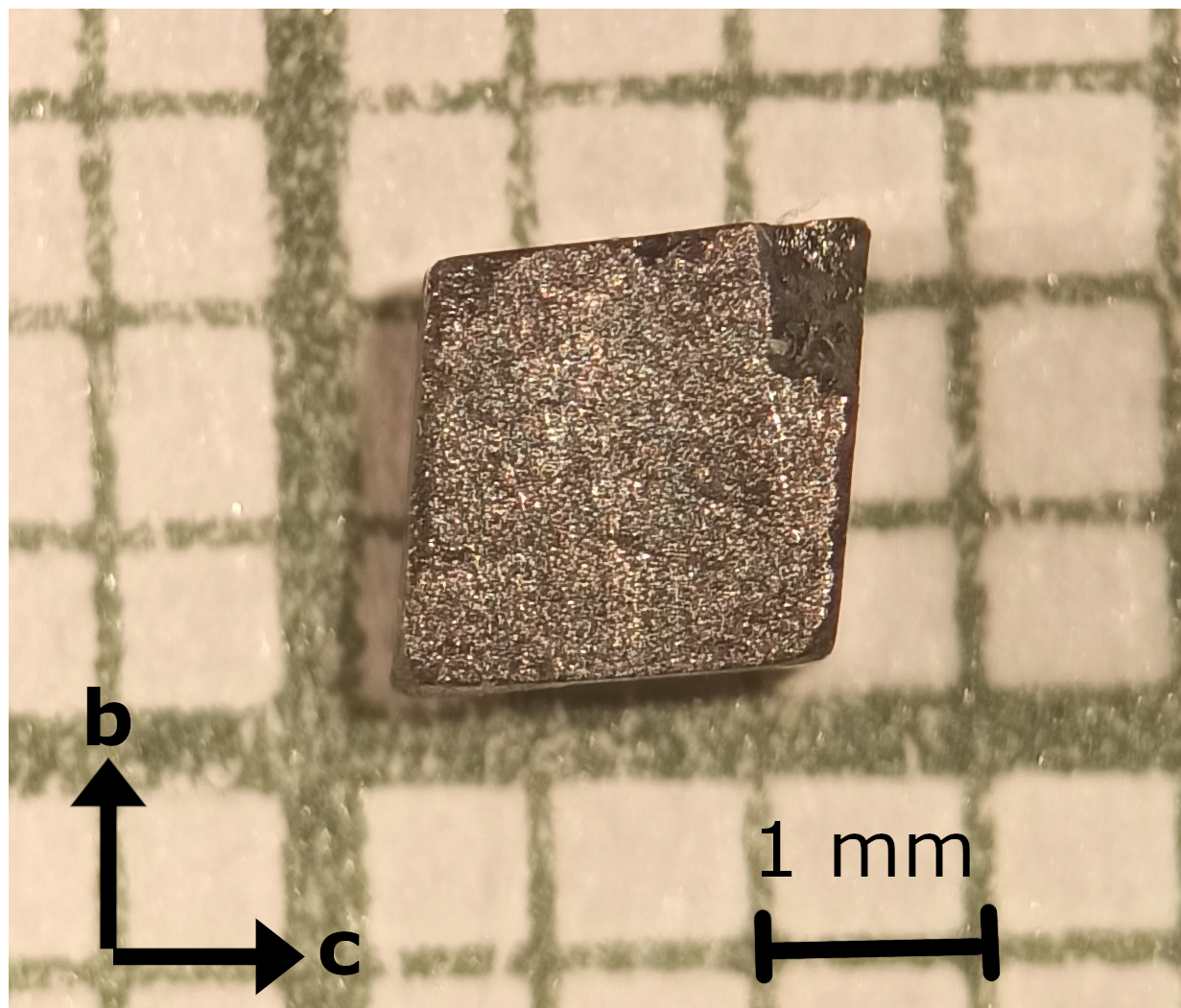}
    \caption{Picture of the \mto\ single crystal used for the magnetisation, thermal expansion and heat capacity measurements.}
    \label{fig:Crystal_Pic}
\end{figure}

\begin{figure}[h]
    \centering
    \includegraphics[width=0.9\columnwidth,clip]{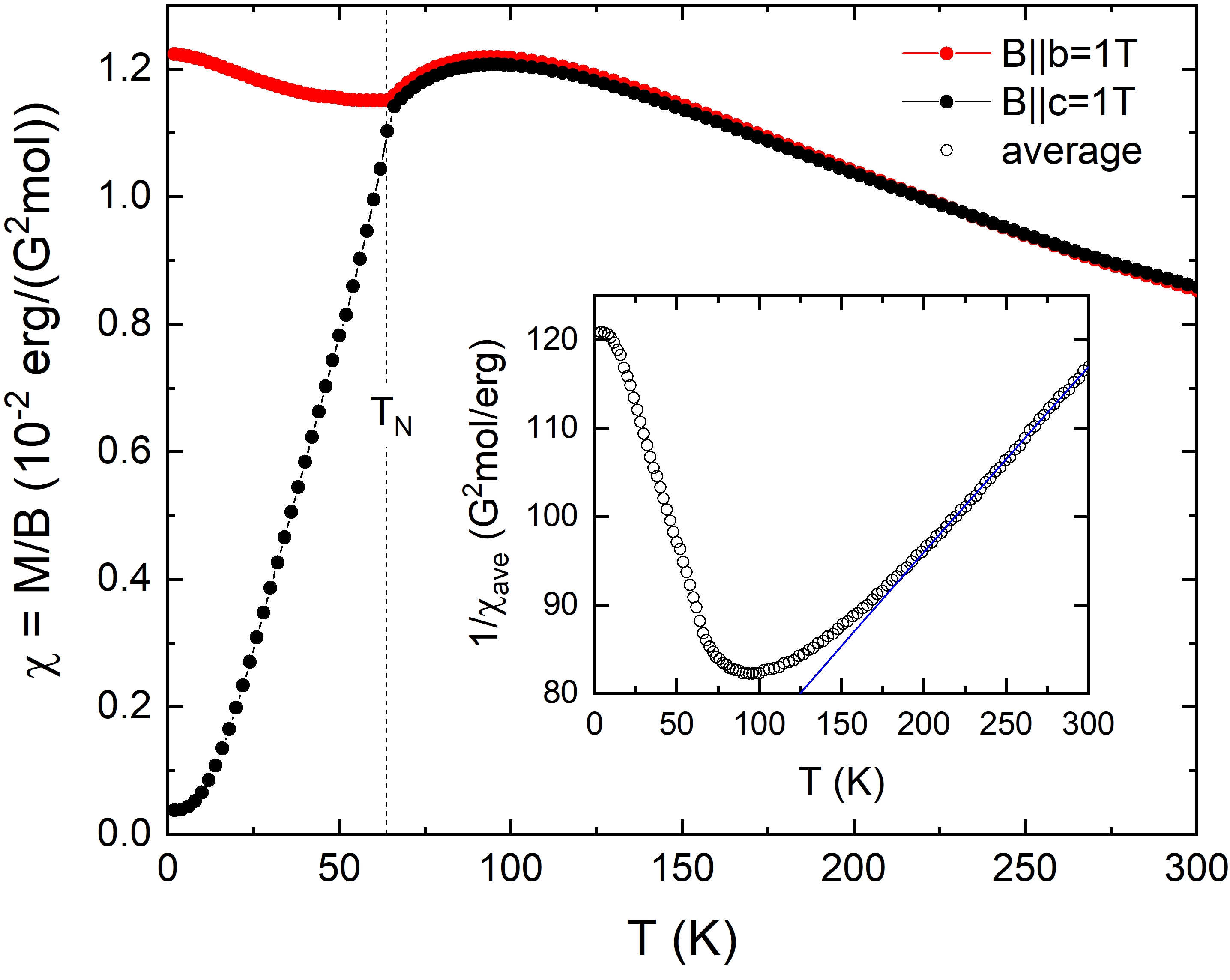}
    \caption{Static magnetic susceptibility, $\chi=M/B$, along the two principal axes, $B||b$ (closed red points) and $B||c$ (closed black points), obtained at $B=1$~T in a zero-field-cooled mode. Dashed vertical line marks the antiferromagnetic ordering temperature. Inset depicts the inverse of the average static susceptibility (open black points) fitted by means of a Curie-Weiss law (solid blue line).}
    \label{fig:static_magnetic_susceptibility}
\end{figure}

\begin{figure}[htb]
    \centering
    \includegraphics[width=0.9\columnwidth,clip]{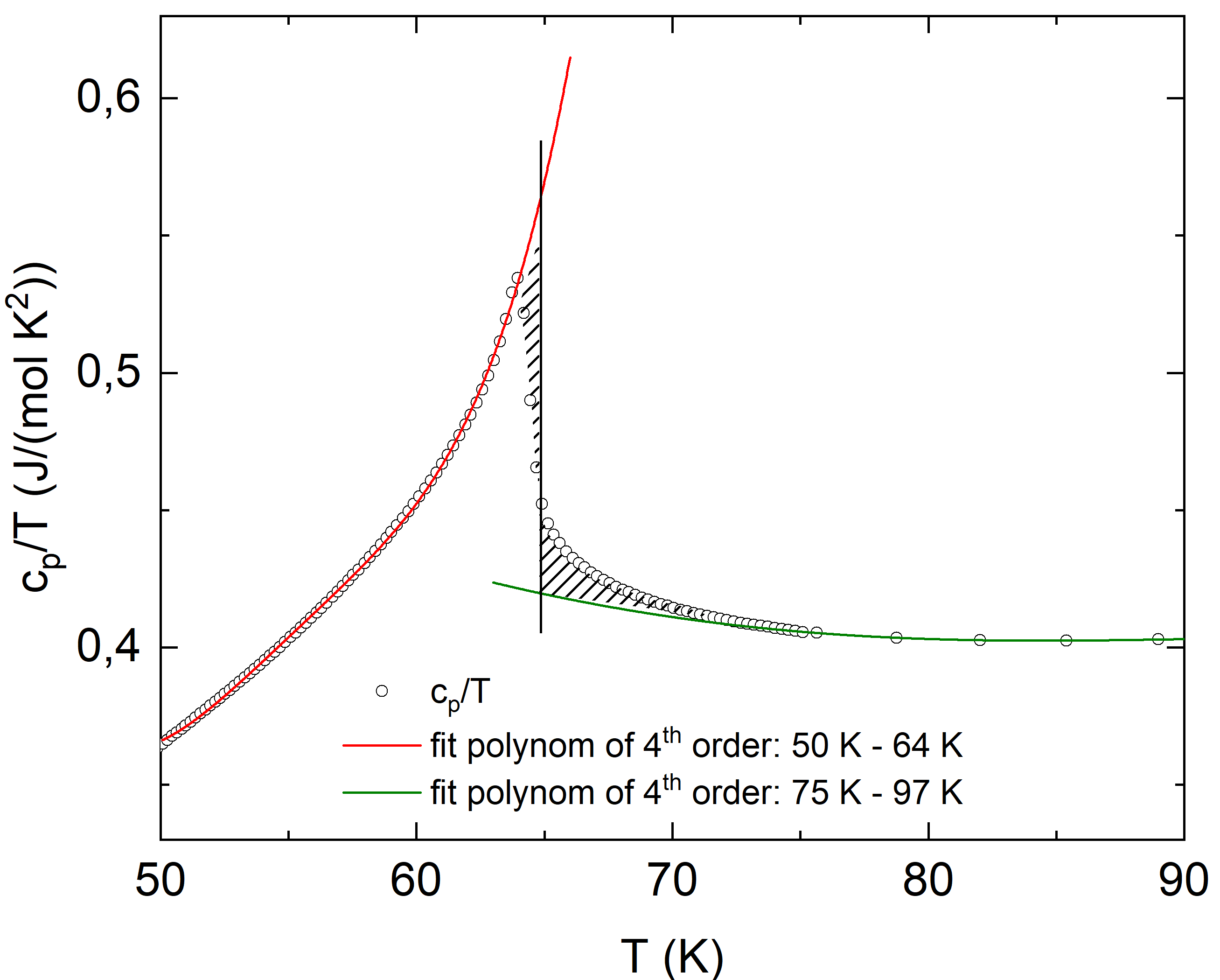}
    \caption{Equal-area construction for the anomaly in $c_p/T$ at \tn, in order to estimate the transition height. Therefore the behaviour of the measured quantity was extrapolated around \tn\ by fitting polynomials of fourth order onto a interval below (red) and above (green) \tn . The exact position of the transition was determined such that the areas (black stripes) below and above \tn\ match in order to ensure entropy and volume conservation. Analogous constructions were applied to obtain the anomalies in \ali .}
    \label{fig:SAC}
\end{figure}

\begin{figure}[h]
    \centering
    \includegraphics[width=0.9\columnwidth,clip]{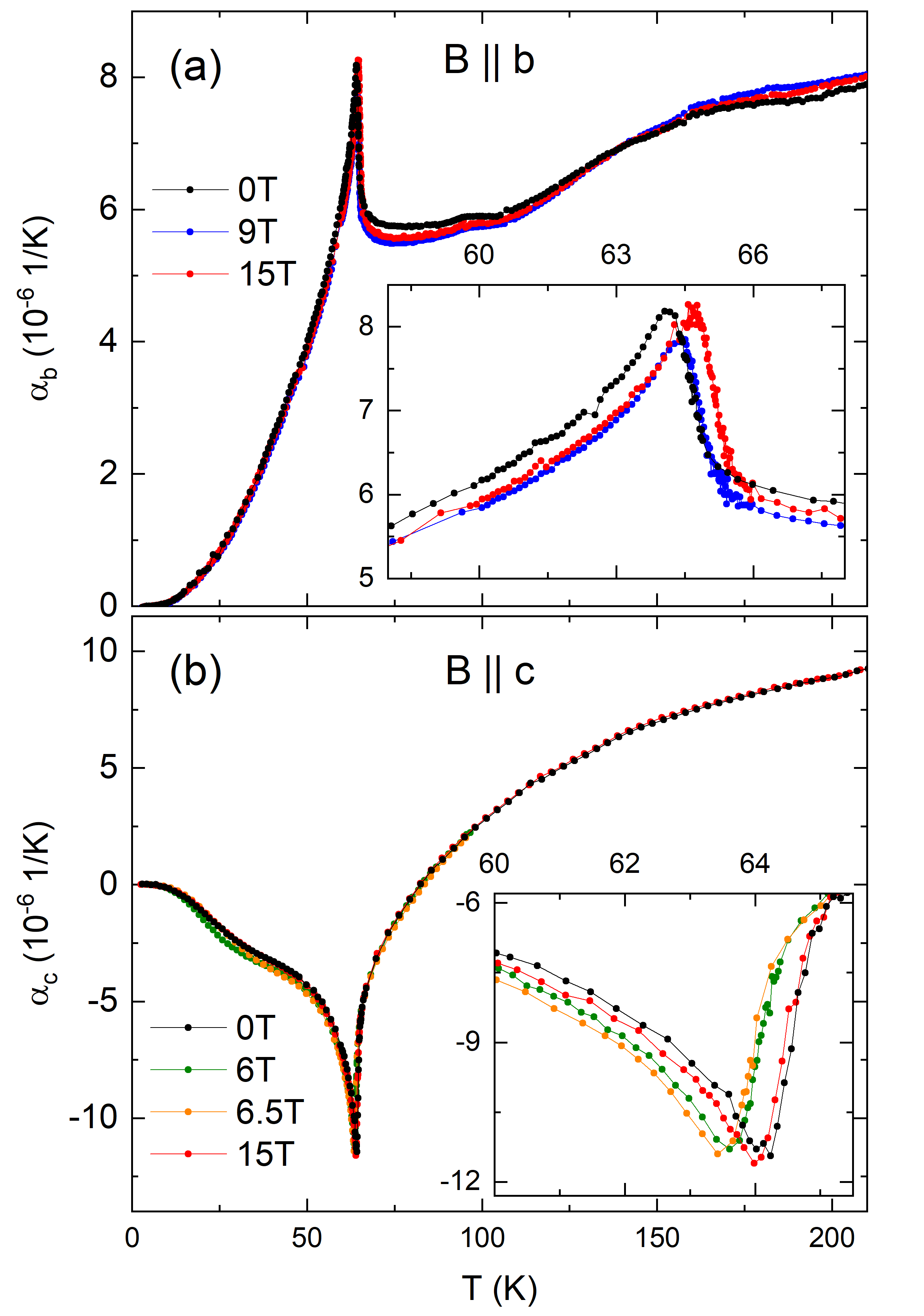}
    \caption{Thermal expansion coefficients $\alpha_i$ $vs.$ temperature for various fields up to $15$~T applied along the (a) $b$-axis and (b) $c$-axis. Insets: Zoomed in temperature region around $T_N$.}
    \label{fig:alpha_in_field}
\end{figure}

\begin{figure}[h]
\centering
\includegraphics[width=0.9\columnwidth,clip]{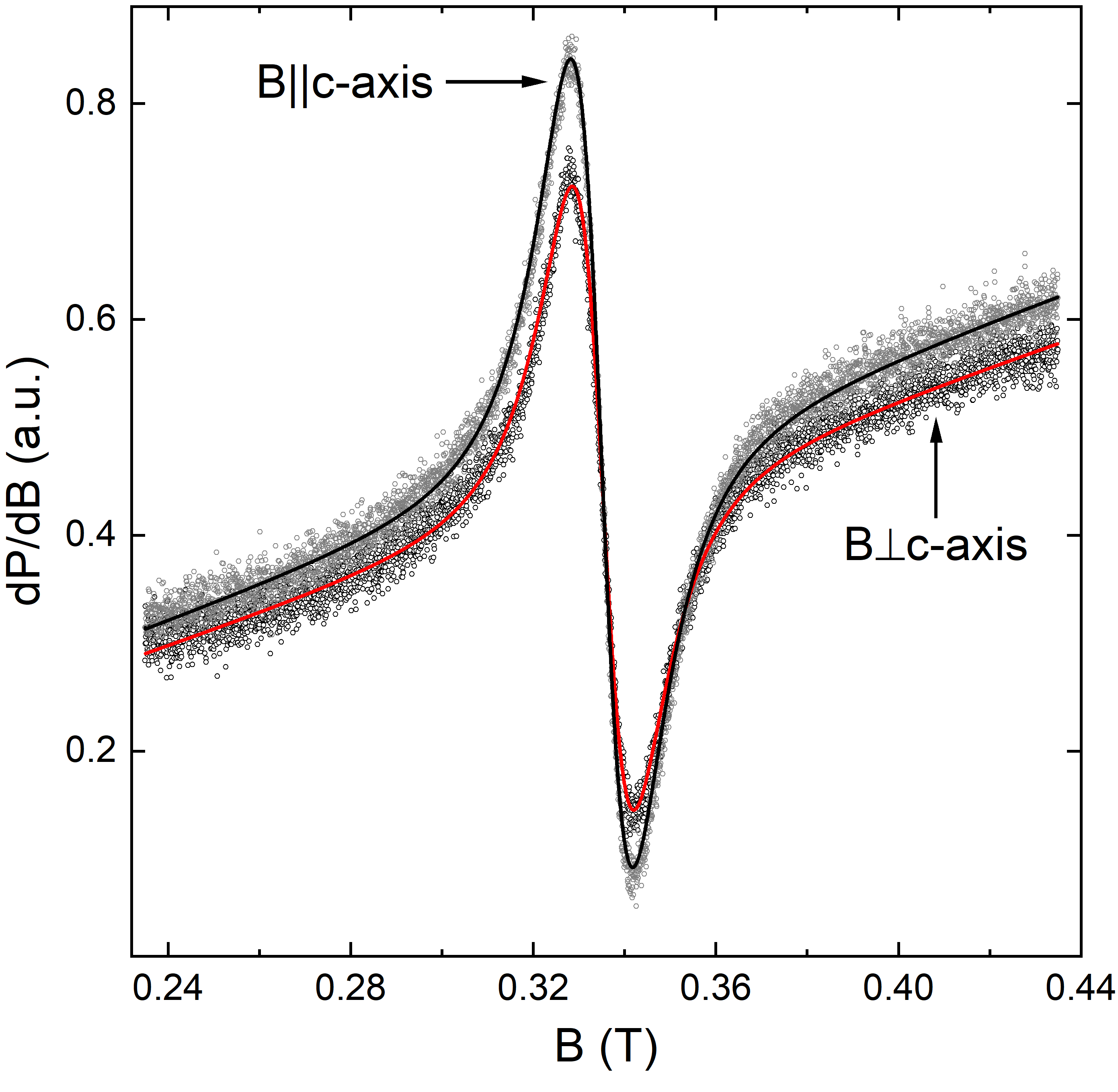}
\caption{Transmission derivative of a paramagnetic resonance feature of \mto\ at X-band frequency of $\nu=9.395$~GHz at room temperature with the external magnetic field applied parallel and perpendicular to the $c$ axis, obtained in a Voigt configuration ($B \perp B_{\mathrm{MW}}$). Black and red solid lines are fitted Lorentzian-peak derivatives with a linear background.}
\label{fig:X-band_RT}
\end{figure}